\documentclass[10pt,conference]{IEEEtran}
\PassOptionsToPackage{hidelinks}{hyperref}

\usepackage[T1]{fontenc}
\usepackage[stretch=10,shrink=10]{microtype}
\usepackage{graphicx}
\usepackage{amsmath,amssymb}
\usepackage{cite}

\usepackage{booktabs}
\usepackage{multirow}
\usepackage{array}
\usepackage{siunitx}
\usepackage[table]{xcolor}
\usepackage{xurl}

\definecolor{qwenC}{HTML}{2A78D6}
\definecolor{devC}{HTML}{EB6834}

\definecolor{glmC}{HTML}{8A6D1F}

\usepackage{tikz}
\usetikzlibrary{positioning,fit,arrows.meta,calc,backgrounds}

\usepackage[most]{tcolorbox}
\newtcolorbox{rqanswer}{breakable, colback=gray!7, colframe=black!55, boxrule=0.6pt,
  arc=1.5pt, left=4pt, right=4pt, top=3pt, bottom=3pt,
  before skip=5pt, after skip=5pt}

\definecolor{findingaccent}{HTML}{2A6B9C}
\newtcolorbox{findingbox}{enhanced, breakable, lines before break=3,
  colback=findingaccent!7, colframe=findingaccent!30, boxrule=0.4pt,
  borderline west={2.5pt}{0pt}{findingaccent}, sharp corners,
  boxsep=0pt, left=5pt, right=5pt, top=2.5pt, bottom=2.5pt,
  before skip=5pt, after skip=9pt}

\usepackage{hyperref}
\hypersetup{
  pdftitle={Pairing Matters: Performance, Failures, and Efficiency in Lightweight Coding Agents},
  pdfauthor={Lu Zhao, Christopher M. Poskitt, David Lo},
  pdfsubject={Harness--model pairing performance, failures, and efficiency in lightweight coding agents},
  pdfkeywords={coding-agent performance, harness-model pairing, token efficiency, trajectory-based failure analysis}
}
\let\originalthebibliography\thebibliography
\renewcommand{\thebibliography}[1]{\originalthebibliography{#1}\interlinepenalty=10000\relax}

\begin{document}
\bstctlcite{IEEEcontrol}
\frenchspacing
\raggedbottom
\clubpenalties 2 10000 0
\widowpenalties 2 10000 0

\title{Pairing Matters: Performance, Failures,\\
and Efficiency in Lightweight Coding Agents}

\author{
\IEEEauthorblockN{
    Lu Zhao\hspace{20pt}
    Christopher M. Poskitt\hspace{20pt}
    David Lo
}
\IEEEauthorblockA{
    \textit{School of Computing and Information Systems},
    \textit{Singapore Management University}, Singapore\\
    lu.zhao.2023@engd.smu.edu.sg, \{cposkitt, davidlo\}@smu.edu.sg
}
}

\maketitle

\begin{abstract}
Self-hosting coding agents is often necessary for software-engineering teams that must keep proprietary code and development traces within controlled infrastructure. Lightweight, open-weight models make such deployment more feasible by reducing hardware requirements, while strong results on repository-level benchmarks demonstrate their potential for demanding coding tasks. Deploying these models as coding agents requires pairing them with an agent harness, making their joint performance a key deployment consideration. Yet existing leaderboards primarily report fixed harness--model configurations rather than controlled comparisons across alternative pairings.
To address this gap, we investigate harness--model interactions through a controlled full-factorial evaluation of six open-source harnesses and three lightweight coding models across 100 SWE-bench Verified tasks, producing 1,800 graded runs under common serving, metering, and grading conditions. For the same model, resolution rates differ by up to 44 percentage points across harnesses, while harness rankings change across models. We apply an evidence-based failure taxonomy to 1,100 unsuccessful runs and find substantial differences in failure patterns across harness--model pairings. A model's dominant failure modes under one harness need not dominate under another. Efficiency is similarly dependent on the pairing: even among model-directed loop harnesses, charged tokens per resolved task vary by more than 20-fold.
These findings make harness--model selection a central deployment decision for lightweight coding agents. Strong performance with one harness or model does not guarantee similar results with another. We translate these findings into empirically grounded practitioner guidance for pairing selection, token-efficiency assessment, and trajectory-informed harness engineering, and release the complete experimental records and analysis artifacts.
\end{abstract}

\begin{IEEEkeywords}coding-agent performance, harness--model pairing, token efficiency, trajectory-based failure analysis\end{IEEEkeywords}

\section{Introduction}\label{sec:background}

\begin{figure*}[t]
  \centering
  \includegraphics[width=\textwidth]{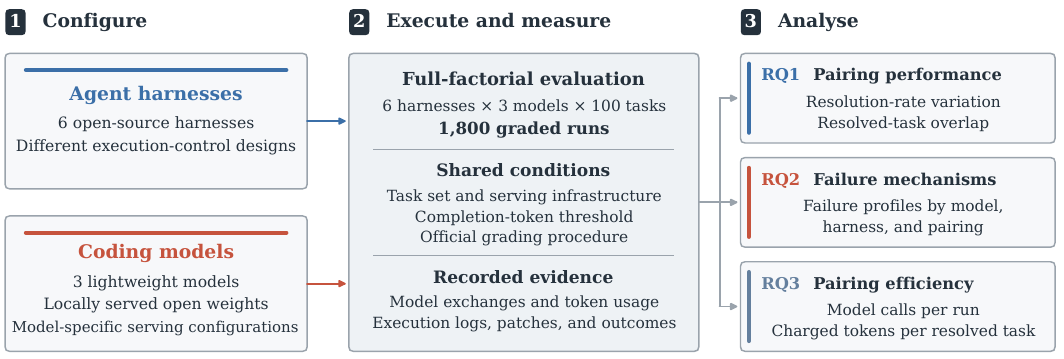}
  \caption{Study design and evidence flow. (1) Configure: select agent harnesses and coding models. (2) Execute and measure: evaluate all 18 pairings on 100 SWE-bench Verified tasks. (3) Analyse: examine pairing performance (RQ1), failure mechanisms (RQ2), and pairing efficiency (RQ3).}
  \label{fig:overview}
\end{figure*}

\begingroup\looseness=1 Confidentiality requirements and the need to retain control over proprietary code motivate local or self-hosted deployment of coding agents~\cite{llamacoderefinement,ipmodernization,codestral2508}. Lightweight,
open-weight models lower the hardware barrier to such deployment, bringing
it within reach of more modest computing resources~\cite{codestral2508}.
Their appeal extends beyond deployability: recent coding-specialized models
demonstrate substantial capability in repository-level issue resolution.
The 24B-parameter Devstral Small reports
53.6\% resolution on SWE-bench Verified with OpenHands~\cite{devstral2507},
while Qwen3-Coder-Next reports 70.6--71.3\% across three harnesses, remaining
competitive with substantially larger models despite activating only 3B of
its 80B parameters~\cite{qwen3codernextreport}. These results position
lightweight models as credible foundations for self-hosted coding agents,
combining deployment flexibility with the capability to tackle demanding
software-engineering tasks.\par\endgroup

Deploying these models as coding agents requires selecting an accompanying
harness, which manages repository context, tool access, execution, validation,
and patch submission~\cite{opencode,openhands,minisweagent,sweagent}. The
deployment decision therefore concerns a harness--model pairing, not either
component alone. Yet existing evaluations offer limited guidance for this
choice. Our audit covered 324 submission records across six tracks of the
official August 2026 SWE-bench archive~\cite{swebenchexperiments}. At least 223 reported results using
proprietary models, compared with five reporting results for off-the-shelf
models with fewer than 100B total parameters paired with established agent
harnesses. These five submissions report fixed
harness--model configurations, providing little systematic evidence of how
changing the harness affects the same lightweight model's performance.
Beyond leaderboards, model reports and harness studies typically evaluate only one or a few fixed harness--model configurations~\cite{devstral2507,qwen3codernextreport,opencode,openhands,minisweagent,sweagent}. Although coding-agent performance is increasingly understood as a property of the joint system, the effects of different pairings remain insufficiently characterized. Existing evidence provides limited insight into how pairing choices affect resolution performance, failure mechanisms, and cost, or how these differences can guide pairing selection and adaptation for specific tasks and resource budgets.

We address this gap through a controlled full-factorial study of three
lightweight coding models and six open-source harnesses. We pair
Qwen3-Coder-Next~\cite{qwen3codernext}, GLM-4.7-Flash~\cite{glm47flash}, and
Devstral-Small~\cite{devstral2507} with OpenCode~\cite{opencode},
OpenHands~\cite{openhands}, mini-SWE-agent~\cite{minisweagent},
SWE-agent~\cite{sweagent}, AutoCodeRover~\cite{autocoderover}, and
Agentless~\cite{agentless}.
Each of the 18 pairings attempts the same 100 SWE-bench Verified tasks,
yielding 1{,}800 graded runs under common serving, metering, and grading
conditions. This design examines how pairing choice shapes resolution
accuracy, resolved-task overlap, failure mechanisms, model-call demand, and
charged tokens per resolved task. Figure~\ref{fig:overview} summarizes the
study, and Section~\ref{sec:design} details the selection criteria and
experimental procedures.

Our results yield three principal insights for deployment. First, harness
selection changes the resolution rate of the same coding model by up to 44
percentage points, and harness rankings do not transfer consistently across
models. Second, failure mechanisms are pairing-dependent: a harness can
attenuate, amplify, or redirect a model's dominant failure tendency. Third,
even among model-directed loop harnesses, charged tokens per resolved task
vary by more than 20-fold. Practitioners should therefore avoid selecting or dismissing a model based on one harness, or a harness based on one model, and consider the tasks each pairing resolves. Execution budgets should account for expenditure on unsuccessful runs rather than assume that lightweight models are token-efficient, while pairing-specific failure evidence should guide targeted harness improvements.

These findings emerge from three complementary analyses, organized around the following research questions:

\begin{itemize}
    \item \textbf{RQ1: Pairing Performance.}\\
    How does task-resolution accuracy vary across harness--model pairings,
    and how do resolved-task sets overlap across models and harnesses?

    \item \textbf{RQ2: Failure Mechanisms.}\\
    Which failure mechanisms occur in unsuccessful runs, and how do failure
    profiles vary across models, harnesses, and pairings?

    \item \textbf{RQ3: Pairing Efficiency.}\\
    How do harness--model pairings differ in model calls per run and charged tokens per resolved task?
\end{itemize}

By addressing these questions, we provide an empirical foundation for understanding harness--model pairing behaviour and informing the deployment of lightweight coding agents. Our contributions are fourfold:

\begin{itemize}
    \item \textbf{A systematic evaluation of lightweight coding-agent pairings.}
    We conduct a full-factorial evaluation of six open-source harnesses and
    three lightweight coding-specialized models under common self-hosted
    serving conditions, characterizing variation in resolution performance,
    resolved-task coverage, and token efficiency.

    \item \textbf{An evidence-based analysis of pairing-specific failures.}
    We develop a taxonomy of five behavioural failure mechanisms and apply
    it to all 1{,}100 unsuccessful runs, enabling comparisons of failure
    profiles across models, harnesses, and individual pairings.

    \item \textbf{Empirically grounded guidance for pairing-aware deployment.}
    We recommend testing alternative harnesses before ruling out a model, and alternative models before ruling out a harness, using both resolution rate and the tasks resolved on the intended workload to guide selection. We also recommend budgeting against resolved tasks, assessing the token efficiency of lightweight models empirically, and using pairing-specific failure evidence to prioritize improvements to progress control, fault localization, and validation.

    \item \textbf{An auditable experimental record.}
    We publicly release records for all 1{,}800 graded runs, including benchmark
    grades, available trajectories, candidate patches, and serving
    telemetry, together with failure labels, configurations, and analysis
    scripts.
\end{itemize}

\begin{table*}[t]
  \centering
  \caption{Execution-control patterns and workflows of the evaluated agent harnesses.}
  \label{tab:scaffolds}
  \definecolor{groupband}{HTML}{E8EEF5}
  \small
  \setlength{\tabcolsep}{6pt}
  \renewcommand{\arraystretch}{1.12}
  \begin{tabular}{@{}
    >{\raggedright\arraybackslash}p{0.30\textwidth}
    >{\raggedright\arraybackslash}p{\dimexpr0.70\textwidth-2\tabcolsep\relax}@{}}
    \toprule
    \textbf{Agent harness} & \textbf{Workflow and evaluated configuration} \\
    \midrule
    \rowcolor{groupband}[0pt][0pt]
    \multicolumn{2}{@{}l@{}}{\textbf{Model-directed interaction loops}} \\
    \addlinespace[2pt]
    OpenCode (OC)~\cite{opencode}
      & Provides structured tools for shell execution, file reading and
        editing, and repository search. \\
    \addlinespace[2pt]
    OpenHands (OH)~\cite{openhands}
      & Uses CodeAct with command execution and file-editing tools in a
        persistent sandbox. \\
    \addlinespace[2pt]
    mini-SWE-agent (mini-SWE)~\cite{minisweagent}
      & Uses a bash-only interface: the model generates shell commands
        and receives their execution output. \\
    \addlinespace[2pt]
    SWE-agent (SA)~\cite{sweagent}
      & Provides an agent--computer interface for repository inspection,
        code editing, shell execution, and patch submission. \\
    \midrule
    \rowcolor{groupband}[0pt][0pt]
    \multicolumn{2}{@{}l@{}}{\textbf{Phased retrieval and repair}} \\
    \addlinespace[2pt]
    AutoCodeRover (ACR)~\cite{autocoderover}
      & Uses iterative search over classes and methods followed by patch
        generation. In our evaluated configuration, generated reproduction scripts can execute, but the
        dedicated regression-test phase is omitted. \\
    \midrule
    \rowcolor{groupband}[0pt][0pt]
    \multicolumn{2}{@{}l@{}}{\textbf{Fixed repair pipeline}} \\
    \addlinespace[2pt]
    Agentless (A)~\cite{agentless}
      & Follows a predefined pipeline of localization, repair, and patch validation, without model-directed command execution. In our evaluation, we retain the localization stages, generate a single candidate patch, and omit internal test-based validation. \\
    \bottomrule
  \end{tabular}
\end{table*}

\section{Study Design}\label{sec:design}

We use a full-factorial design in which each selected coding model is evaluated with every harness. The following subsections describe harness and model selection, benchmark tasks, execution controls, and evaluation procedures.

\subsection{Agent Harness Selection}\label{sec:scaffolds}

We identified candidate harnesses through SWE-bench submissions, published
harness papers, and GitHub searches. Eligibility required a public,
version-pinnable implementation supporting headless execution and model
substitution through our locally served OpenAI-compatible endpoint.
Where needed, lightweight adapters handled harness invocation and
connection to our locally served models while preserving each harness's
existing interaction workflow.

Among the harnesses meeting these criteria, we selected six to balance research or practitioner adoption with diversity in execution control. At our July 2026 selection cutoff, their repositories had approximately 2{,}000 to 200{,}000 GitHub stars; we use these dated counts as indicators of adoption, not quality. Prior work distinguishes workflow-based repair agents from agents with model-directed execution~\cite{traceability}. Drawing on this distinction and the documented designs of the evaluated systems, we group their execution control into model-directed interaction loops, phased retrieval and repair, and a fixed repair pipeline~\cite{autocoderover,agentless}. The selection prioritizes practically relevant and contrasting designs rather than exhaustive coverage. Table~\ref{tab:scaffolds} summarizes their execution-control patterns and workflows.

\smallskip\noindent\textbf{Model-directed interaction loops.}
OpenCode~\cite{opencode}, OpenHands~\cite{openhands}, mini-SWE-agent~\cite{minisweagent}, and SWE-agent~\cite{sweagent} follow an action--observation loop: the model selects actions using task instructions and harness-provided context, while the harness executes them and returns observations. This cycle continues until completion or a resource limit is reached. Their tool interfaces and execution environments differ, as summarized in Table~\ref{tab:scaffolds}.

\smallskip\noindent\textbf{Phased retrieval and repair.}
AutoCodeRover uses structure-aware retrieval followed by patch generation~\cite{autocoderover}. The model directs searches over program elements, such as classes and methods, while the harness controls phase progression and tool availability. In our evaluated configuration, model-generated reproduction scripts can execute, but the dedicated regression-test phase is omitted.

\smallskip\noindent\textbf{Fixed repair pipeline.}
Our evaluated Agentless~\cite{agentless} configuration follows predefined localization stages and generates a single candidate patch. The model supplies stage outputs, while the harness controls progression and termination without a model-directed command interface. Internal test-based patch validation is omitted; the resulting patch is assessed by the common SWE-bench grader.

\subsection{Model Selection}\label{sec:models}

We considered coding-specialized open-weight models released during the year preceding our July 2026 selection. Eligibility required 10--100B total parameters and public checkpoints that could be served on a single H100-class GPU under the evaluated serving configurations. Among models meeting these criteria, we selected three based on reported SWE-bench Verified performance, practitioner adoption indicated by Hugging Face download counts recorded at selection, and diversity in architecture, training emphasis, reasoning mode, and tool-use protocol. The selection prioritizes contrasting designs relevant to self-hosted deployment rather than aiming to represent all eligible models.

\noindent\textbf{Qwen3-Coder-Next (Qwen3-Coder).}
This sparse mixture-of-experts model has 80B total and approximately 3B active parameters. Its training emphasizes long-horizon
coding-agent tasks, tool use, and recovery from execution
failures~\cite{qwen3codernext}. 

\smallskip\noindent\textbf{Devstral-Small (Devstral-S).}
This dense 24B model was developed by Mistral AI and All Hands AI for
agentic software engineering~\cite{devstral2507}. Its development with the OpenHands maintainers motivates comparing its performance under OpenHands with that under other harnesses.

\smallskip\noindent\textbf{GLM-4.7-Flash (GLM-4.7-F).}
This mixture-of-experts model has 30B total and approximately 3B active parameters and supports explicit reasoning~\cite{glm47flash}. We evaluate it with thinking enabled throughout the study, in contrast to the non-thinking configurations of the other two models.

\begingroup
\microtypesetup{expansion=false}
Together, the three models span dense and sparse architectures and differ in agentic training, reasoning modes, and tool-use protocols. These contrasts allow us to examine harness--model pairing behaviour across different model designs, while keeping conclusions scoped to the evaluated models and configurations.
\par
\endgroup

\subsection{Benchmark Selection}\label{sec:dataset}

To keep the evaluation of all 18 harness--model pairings within our computational budget, we use a fixed subset of 100 real-world software-engineering tasks from the 500-task SWE-bench Verified benchmark~\cite{swebench,swebenchverified}. Task selection emphasizes repository coverage while incorporating the benchmark's official difficulty categories. We use fixed-seed sampling: initial draws balance allocations across repositories, while subsequent draws allocate quotas across repository--difficulty strata in proportion to the full benchmark. The resulting subset therefore gives smaller repositories greater representation rather than reproducing the full benchmark's repository proportions. Tasks with grading defects identified during screening or execution were excluded and, where required, replaced using documented seeded procedures. Table~\ref{tab:tasks} summarizes the final subset's repository and difficulty composition.
\begin{table}[t]
  \centering
  \caption{Distribution of tasks: SWE-bench Verified vs.\ our
  subset, by repository and by the official difficulty annotation.}
  \label{tab:tasks}
  \small
  \setlength{\tabcolsep}{6pt}
  \begin{tabular}{@{}lrr@{}}
    \toprule
    & Verified & Ours \\
    \midrule
    \multicolumn{3}{@{}l}{\textit{Repository}} \\
    astropy/astropy           &  22 &  7 \\
    django/django             & 231 & 33 \\
    matplotlib/matplotlib     &  34 &  9 \\
    mwaskom/seaborn           &   2 &  2 \\
    pallets/flask             &   1 &  1 \\
    psf/requests              &   8 &  3 \\
    pydata/xarray             &  22 &  8 \\
    pylint-dev/pylint         &  10 &  4 \\
    pytest-dev/pytest         &  19 &  7 \\
    scikit-learn/scikit-learn &  32 &  8 \\
    sphinx-doc/sphinx         &  44 &  7 \\
    sympy/sympy               &  75 & 11 \\
    \midrule
    \multicolumn{3}{@{}l}{\textit{Difficulty (official annotation)}} \\
    $<$15 min fix             & 194 & 35 \\
    15 min -- 1 hour          & 261 & 56 \\
    1 -- 4 hours              &  42 &  9 \\
    $>$4 hours                &   3 &  0 \\
    \midrule
    Total                     & 500 & 100 \\
    \bottomrule
  \end{tabular}
\end{table}

\subsection{Experimental Runtime}\label{sec:harness}

We developed a common experimental runtime that prepares task repositories, invokes harnesses, serves models, and collects outputs across all 18 pairings. Harnesses run
headlessly, with Docker-based environments used for task execution.
Available workspaces and run outputs are retained regardless of
termination status. Section~\ref{sec:grading-metrics} describes the recorded
artifacts, efficiency metrics, and grading procedure.

\smallskip\noindent\textbf{Execution controls.}
All pairings shared the task set, serving infrastructure, grading procedure, and a cumulative threshold of 262,144 completion tokens per run. Harness-specific limits and timeouts provided additional execution bounds. Starting from native configurations, we adjusted selected limits in response to prolonged runs during evaluation to keep execution computationally tractable. These additional bounds are not directly equivalent across harnesses; our results therefore characterize the evaluated deployment configurations. The released artifact documents the exact settings.

\smallskip\noindent\textbf{Model serving and hardware.}
We serve models sequentially through vLLM, keeping each model's serving configuration fixed across harnesses. All harnesses use an OpenAI-compatible interface, with model requests and token usage recorded through the common metering infrastructure. Prefix caching is enabled, with model-specific parsers configured for tool-calling requests. Inference runs on a single NVIDIA H100 NVL GPU, while harnesses execute on a separate CPU host connected through an SSH tunnel.

\subsection{Grading and Metrics}\label{sec:grading-metrics}

We use a harness-independent measurement protocol to compare resolution performance, trajectory length, and token expenditure across pairings. The common proxy records model exchanges and token usage reported by the serving engine. Execution logs, run metadata, candidate patches, and benchmark grades provide complementary evidence for outcome assessment and failure analysis.

\smallskip\noindent\textbf{Run and Trajectory.}
A run is one attempt by a harness--model pairing on one task. Its trajectory is the sequence of model requests and responses, including tool calls and observations represented in those exchanges. For each run, we associate the available trajectory with its execution logs, run metadata, candidate patch, and benchmark grade. The run metadata records termination status, wall-clock duration, and patch availability. Candidate patches are retained for benchmark grading and failure analysis.

The proxy uses a common recording format across harnesses. Its metering records include prompt and completion token counts and, where available, timestamps, latency, time to first token, cached-token counts, and finish reasons.

\smallskip\noindent\textbf{Trajectory Length in Iterations.}
A run's trajectory length is the total number of model calls made during that run. Each model request and its corresponding response constitute one iteration. When harness-initiated retries or sub-agent calls generate separate model calls, each such call is counted as a separate, additional iteration.

\smallskip\noindent\textbf{Token Cost.}
We calculate a run's charged-token cost by summing the prompt and
completion tokens reported by the serving engine across all model calls.

For execution control, we set a shared generation budget of 256K (262{,}144) completion
tokens per run, accumulated across model calls and excluding prompt
tokens. This study-specific threshold accommodates the completion-token
usage observed in most runs while providing a common safeguard against
excessive generation during prolonged or repetitive execution.

\smallskip\noindent\textbf{Grading and Resolution.}
Before grading, we filter candidate patches to remove edits to tests, selected configuration files, and auxiliary files such as documentation and temporary scripts. We use the resolution outcomes reported by the official SWE-bench Docker grader~\cite{swebench}. Runs with missing or empty patches are counted as unsuccessful. A pairing's resolution rate is the proportion of its 100 tasks resolved.

\section{RQ1: Pairing Performance}\label{sec:rq1} \begin{table}[t]
  \centering
  \caption{Tasks resolved (of 100) per harness--model pairing.}
  \label{tab:grid}
  \small
  \setlength{\tabcolsep}{0pt}
  \begin{tabular}{@{}l@{\hspace{8pt}} c @{\hspace{10pt}} c @{\hspace{10pt}} c@{}}
  \toprule
  Harness & {\footnotesize\textbf{Qwen3-Coder}} & {\footnotesize\textbf{GLM-4.7-F}} & {\footnotesize\textbf{Devstral-S}} \\
  \midrule
  OpenCode       & 61 & \textbf{51} & 44 \\
  OpenHands      & \textbf{63} & 42 & \textbf{54} \\
  mini-SWE-agent & 57 & 49 & 22 \\
  SWE-agent      & 22 & \textbf{51} & 31 \\
  AutoCodeRover  & 19 & 28 & 27 \\
  Agentless      & 39 & 23 & 17 \\
  \midrule
  \emph{Any harness (union)} & \textbf{79} & \textbf{62} & \textbf{64} \\
  \bottomrule
  \end{tabular}
\end{table}

Table~\ref{tab:grid} summarizes resolution performance across the 18 harness--model pairings on our 100-task subset. For each model, the \emph{Any harness (union)} row counts tasks resolved by at least one of the six harnesses. This combined coverage exceeds that of the strongest individual pairing for all three models. For Qwen3-Coder, the union covers 79 tasks, compared with 63 resolved by OpenHands, a gap of 16 tasks. For GLM-4.7-F, the union covers 62 tasks, while OpenCode and SWE-agent tie at 51, leaving a gap of 11. For Devstral-S, the union covers 64 tasks, compared with 54 resolved by OpenHands, a gap of 10.\par Even the best-performing harness for each model therefore leaves a substantial number of tasks unresolved, although the same model resolves them with other harnesses. These results show that performance measured under one harness does not fully capture a model's observed task coverage across harnesses, supporting the joint evaluation of model and harness choices.\begin{findingbox}
\textbf{Finding 1.} For the same model, resolution rates differ by 28--44 percentage points across harnesses. A harness that leads with one model does not necessarily lead with another. Even the best-performing harness for each model resolves 10--16 fewer tasks than the combined coverage across all six harnesses.
\end{findingbox}

\begin{figure}[t]
  \centering
  \includegraphics[width=0.98\columnwidth]{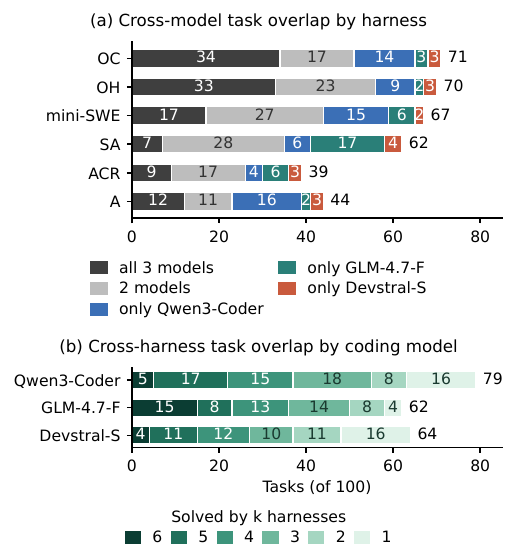}
  \caption{Resolved-task overlap across harness--model pairings: (a) overlap across models for each harness; (b) overlap across harnesses for each model.}
  \label{fig:anchors}
\end{figure}

\begin{table*}[!t]
\caption{Failure taxonomy for the 1{,}100 unsuccessful runs. Reviewers apply the criteria from top to bottom and assign the first matching label. Diagnostic indicators support evidence assessment rather than determine labels automatically.}
\label{tab:ftypes}
\centering\small
\setlength{\tabcolsep}{5pt}
\renewcommand{\arraystretch}{1.18}
\begin{tabular}{@{}
  >{\raggedright\arraybackslash}p{0.20\textwidth}
  >{\raggedright\arraybackslash}p{0.28\textwidth}
  >{\raggedright\arraybackslash}p{\dimexpr0.52\textwidth-4\tabcolsep\relax}@{}}
\toprule
\textbf{Failure type} & \textbf{Diagnostic indicators}
  & \textbf{Assignment criterion} \\
\midrule
\textbf{Exchange breakdown}
  & No model traffic; unanswered or unexecutable tool exchange;
    truncated final exchange; runtime or sandbox failure;
    completed patch lost or invalidated by the harness.
  & An identified, unrecovered interaction or environment failure prevents
    repair completion or valid submission. \\
\midrule
\textbf{No-progress loop or budget exhaustion}
  & Repair remains unfinished when execution stalls, repeats ineffective
    actions, or reaches a step/token limit.
  & The run ends before the repair attempt is finalized because execution
    stalls, repeats ineffective actions, or exhausts its step/token budget. \\
\midrule
\textbf{Fault mislocalization}
  & No shared files between the actual submitted patch and the
    benchmark's gold patch.
  & The submitted patch modifies none of the files modified by the gold
    patch. \\
\midrule
\textbf{Insufficient validation}
  & Success claim about the submitted repair; absent or inadequate
    checks of the required behaviour; tests altered to accommodate
    the patch; relevant contradictory results ignored or misreported.
  & The agent maintains that the submitted repair is successful, but
    the checks and results available during the run do not adequately
    support that claim. \\
\midrule
\textbf{Wrong or incomplete fix}
  & Finalized repair attempt; file-level overlap between the submitted
    and gold patches; failed reference tests.
  & The repair attempt is finalized, but the submitted patch fails
    the reference tests. The patch modifies at least one file also
    changed by the gold patch, and none of the preceding criteria applies. \\
\bottomrule
\end{tabular}
\end{table*}

Figure~\ref{fig:anchors}(a) compares the task sets resolved by each harness across the three models. OpenCode has the largest union, covering 71 tasks, and the largest shared core, with 34 tasks resolved under all three models. OpenHands follows closely, with a union of 70 tasks and a shared core of 33. SWE-agent shows less overlap across models: only 7 tasks in its 62-task union are resolved under all three models, while 17 are resolved exclusively with GLM-4.7-F. Model choice therefore affects not only how many tasks a harness resolves, but also which tasks it resolves.\par Figure~\ref{fig:anchors}(b) compares task coverage across harnesses for each model. Qwen3-Coder achieves the widest combined coverage, with 79 tasks resolved by at least one harness, but only 5 by all six. GLM-4.7-F covers fewer tasks overall, at 62, yet has the largest shared core, with 15 tasks resolved by every harness. Devstral-S covers 64 tasks, including 16 resolved by exactly one harness, indicating that part of its observed coverage depends on the harness selected. These comparisons distinguish breadth of observed coverage from consistency across harnesses: a larger union does not necessarily imply a larger shared core.\par Taken together, these results show comparatively strong consistency in resolved-task coverage across models for OpenCode and OpenHands, and greater consistency in successful outcomes across harnesses for GLM-4.7-F. These tendencies, however, do not imply universal superiority. A model's relative performance depends on the harness, just as a harness's relative performance depends on the model. Component-level strengths can therefore guide selection, but cannot replace evaluation of the intended pairing.

\begin{findingbox}
\textbf{Finding 2.} Models and harnesses have distinct strengths in task
coverage and resolved-task overlap, but these do not transfer uniformly
across pairings. OpenCode and OpenHands share 34 and 33 resolved tasks
across all three models, versus 7 for SWE-agent. Across all six harnesses,
GLM-4.7-F shares 15 tasks versus Qwen3-Coder's 5, despite Qwen3-Coder's
broader combined coverage of 79 versus 62 tasks. Component-level strengths
can guide selection, but the model and harness need to be evaluated jointly.
\end{findingbox}

\section{RQ2: Failure Mechanisms}\label{sec:rq2}

\smallskip\noindent\textbf{Failure taxonomy.}
Prior work examines software-engineering agent trajectories through process-centric analysis of phase transitions and inefficiency patterns~\cite{processcentric}, execution-error analysis~\cite{chenbeyondfinal}, thought--action--result patterns~\cite{bouzeniatrajectories}, and telemetry-grounded diagnosis~\cite{zhaodebuggers}. Building on these approaches, we develop a common run-level taxonomy for comparing pairings, with an emphasis on harness--model interactions, observable execution behaviour, and repair outcomes. We explicitly distinguish exchange breakdowns, where an unrecovered interaction or environment failure prevents repair completion or valid submission, from failures involving execution progress, fault localization, validation, or repair correctness. The categories therefore capture both how the model and harness interact during execution and how the repair attempt progresses and ultimately ends. Table~\ref{tab:ftypes} defines five mutually exclusive behavioural failure types. Criteria are applied from top to bottom, and each run receives the first matching label.

We applied this taxonomy to all 1{,}100 unsuccessful runs through a uniform evidence-based review procedure. Automated scripts extracted diagnostic indicators from proxy records, execution trajectories, and submitted patches. For each run, these indicators were assembled into an evidence pack containing the task description, condensed trajectory, final exchange, and patch text. The indicators supported interpretation of the recorded behaviour rather than determining the label automatically.

Two model reviewers, Claude Sonnet 5 and Claude Haiku 4.5, assessed each run through blinded independent review using the same rubric and the ordered criteria in Table~\ref{tab:ftypes}. During the initial assessment, each reviewer was blinded to existing labels and the other reviewer's ratings. Cases with insufficient evidence underwent evidence enrichment and reassessment before agreement was calculated.

After evidence enrichment and completion of all ratings, the reviewers agreed on 784 of 1{,}100 runs (Cohen's $\kappa = 0.600$). We used unweighted $\kappa$ to account for agreement expected from each reviewer's category frequencies~\cite{cohen1960}. The 316 disagreements underwent a separate adjudication step using Claude Sonnet 5 in fresh contexts. The adjudicator examined both ratings, their rationales, and the recorded evidence, applying the same ordered criteria rather than automatically accepting either reviewer's label. This step produced a final label for all 316 disputed runs.

As an additional quality-control step, Claude Sonnet 5 re-examined 48 cases with matching reviewer ratings in fresh contexts, checking the agreed labels against the recorded evidence and rubric. Of these, 47 labels were retained and one was reassigned to a different failure category. The released artifact documents the reviewer identifiers, prompts, individual ratings, adjudication decisions, and the spot-check correction.

\begin{figure*}[!t]
\centering
\includegraphics[width=0.96\textwidth]{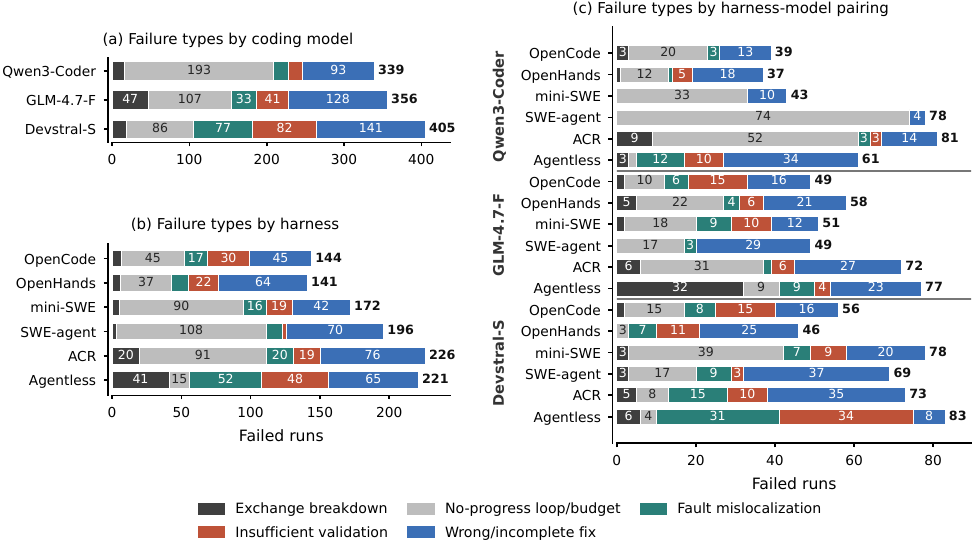}
\caption{Distribution of five failure types across 1{,}100 unsuccessful
runs, aggregated by (a) coding model, (b) agent harness, and (c)
harness--model pairing. Segment widths indicate failure counts; bold
numbers at bar ends indicate total unsuccessful runs.}
\label{fig:pairingcauses}
\end{figure*}
\smallskip\noindent\textbf{Model-level failure profiles.}
Figure~\ref{fig:pairingcauses} summarizes the five failure mechanisms across the 1{,}100 unsuccessful runs. No-progress loops or budget exhaustion form the largest category, with 386 failures (35.1\%), followed by wrong or incomplete fixes with 362 (32.9\%). The remaining categories comprise insufficient validation with 141 (12.8\%), fault mislocalization with 129 (11.7\%), and exchange breakdowns with 82 (7.5\%). Figure~\ref{fig:pairingcauses}(a) aggregates failures by coding model, Figure~\ref{fig:pairingcauses}(b) by agent harness, and Figure~\ref{fig:pairingcauses}(c) by individual pairing.

The model-level aggregates reveal differences in both failure frequency and composition.

Qwen3-Coder exhibits the clearest concentration in a single mechanism. No-progress loops or budget exhaustion account for 193 of its 339 unsuccessful runs (57\%). Exchange breakdown, fault mislocalization, and insufficient validation are comparatively uncommon, contributing 16, 19, and 18 failures, respectively. Although Qwen3-Coder achieves the highest overall resolution rate among the evaluated models, more than half of its unsuccessful runs end without a finalized repair attempt because execution stalls, repeats ineffective actions, or exhausts its step or token budget. Its strong aggregate performance therefore coexists with a pronounced progress-related failure tendency.

Devstral-S shows a contrasting pattern. No-progress loops or budget exhaustion account for 86 of its 405 failures (21\%), while insufficient validation and fault mislocalization account for 82 (20\%) and 77 (19\%), respectively. Wrong or incomplete fixes form its largest category, with 141 failures (35\%). Compared with Qwen3-Coder, its failure profile is less concentrated in progress-related failures, with comparatively greater shares involving localization and validation.

GLM-4.7-F's largest category is wrong or incomplete fix, with 128 of its 356 failures (36\%), followed by no-progress loops or budget exhaustion with 107 (30\%). Exchange breakdowns, insufficient validation, and fault mislocalization account for 47 (13\%), 41 (12\%), and 33 (9\%), respectively. The two largest categories together account for 66\% of its unsuccessful runs.

Beyond resolution accuracy, these failure profiles suggest different priorities for harness engineering. Qwen3-Coder's concentration in no-progress loops or budget exhaustion suggests investigating stronger progress control. Devstral-S's comparatively prominent fault-mislocalization and insufficient-validation failures suggest investigating support for locating faults and checking repairs. For GLM-4.7-F, wrong or incomplete fixes and loop/budget failures together account for two-thirds of unsuccessful runs, suggesting attention to both repair correctness and execution progress.

\begin{findingbox}
\textbf{Finding 3.} Coding models exhibit distinct failure profiles that remain visible across harnesses. Qwen3-Coder is dominated by no-progress loops or budget exhaustion (57\%). Devstral-S shows a more dispersed profile spanning wrong or incomplete fixes (35\%), loops (21\%), insufficient validation (20\%), and fault mislocalization (19\%). GLM-4.7-F is led by wrong or incomplete fixes (36\%) and loops (30\%). These model-level tendencies recur across harnesses, although their severity varies by pairing.
\end{findingbox}

\smallskip\noindent\textbf{Agent harness and pairing effects.}
Figure~\ref{fig:pairingcauses}(b) summarizes each harness's failure profile across models, whereas Figure~\ref{fig:pairingcauses}(c) separates the individual pairings. Comparing these pairings shows why an aggregate failure tendency cannot be attributed to either component alone.

With Qwen3-Coder, no-progress loops or budget exhaustion account for 95\% of unsuccessful SWE-agent runs, compared with 64\% under AutoCodeRover, 51\% under OpenCode, and 32\% under OpenHands. The same mechanism therefore dominates this model's failures under some harnesses but accounts for a much smaller share under others. Qwen3-Coder's aggregate progress-related failure tendency does not characterize all its pairings equally.

\begin{table}[t]
\centering
\caption{Proxy-metered iterations and token expenditure by
harness--model pairing.}
\label{tab:cost}
\begin{minipage}{0.94\linewidth}
\centering
  \small
  \setlength{\tabcolsep}{2pt}
  \renewcommand{\arraystretch}{1.20}
  \begin{tabular}{@{}>{\raggedright\arraybackslash}p{0.30\columnwidth}
    >{\centering\arraybackslash}p{\dimexpr0.70\columnwidth/6-2\tabcolsep\relax}
    S[table-format=2.1,table-column-width=\dimexpr0.70\columnwidth/6-2\tabcolsep\relax]
    >{\centering\arraybackslash}p{\dimexpr0.70\columnwidth/6-2\tabcolsep\relax}
    S[table-format=2.1,table-column-width=\dimexpr0.70\columnwidth/6-2\tabcolsep\relax]
    >{\centering\arraybackslash}p{\dimexpr0.70\columnwidth/6-2\tabcolsep\relax}
    S[table-format=2.1,table-column-width=\dimexpr0.70\columnwidth/6-2\tabcolsep\relax]@{}}
  \toprule
   & \multicolumn{2}{c}{\footnotesize\textbf{Qwen3-Coder}}
   & \multicolumn{2}{c}{\footnotesize\textbf{GLM-4.7-F}}
   & \multicolumn{2}{c}{\footnotesize\textbf{Devstral-S}} \\
  \cmidrule(lr){2-3}\cmidrule(lr){4-5}\cmidrule(lr){6-7}
  Harness & \multicolumn{1}{c}{\footnotesize iters} & \multicolumn{1}{c}{\footnotesize tok/slv}
          & \multicolumn{1}{c}{\footnotesize iters} & \multicolumn{1}{c}{\footnotesize tok/slv}
          & \multicolumn{1}{c}{\footnotesize iters} & \multicolumn{1}{c}{\footnotesize tok/slv} \\
  \midrule
  OpenCode       & 110 & 26.3 & 65  & 6.9  & 23 & 6.9 \\
  OpenHands      & 119 & 15.6 & 57  & 5.1  & 39 & 4.3 \\
  mini-SWE-agent & 128 & 9.5  & 92  & 6.2  & 62 & 9.9 \\
  SWE-agent      & 169 & 87.8 & 147 & 11.0 & 25 & 6.7 \\
  AutoCodeRover  & 50  & 4.8  & 86  & 1.2 & 20 & 1.1 \\
  Agentless      & 4   & 0.1 & 4   & 0.2  & 4  & 0.2 \\
  \midrule
  \emph{Model aggregate}
    & {\raisebox{0.15ex}{---}} & 19.8
    & {\raisebox{0.15ex}{---}} & 6.0
    & {\raisebox{0.15ex}{---}} & 5.1 \\
  \bottomrule
  \end{tabular}
  \par\vspace{4pt}
  \begin{minipage}{\linewidth}
  \footnotesize\raggedright
  \textit{Notes.} \textit{iters}: median model calls per run, rounded to the nearest integer.\par
  \textit{tok/slv}: total charged tokens (prompt + completion) across
  successful and unsuccessful runs, divided by resolved tasks, in millions.\par
  \textit{Model aggregate}: total tokens across harnesses divided by
  the sum of their resolved-task counts; pooled iteration medians are not reported.
  \end{minipage}
\end{minipage}
\end{table}

\begin{figure}[t]
\centering
\includegraphics[width=0.90\linewidth]{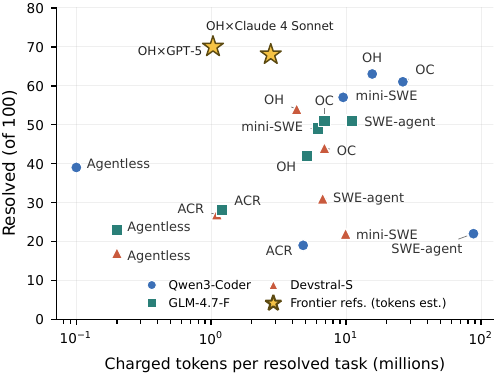}
\caption{Task resolution and token efficiency across harness--model pairings. Token usage is proxy-metered for the 18 evaluated pairings. Stars denote OpenHands-based frontier references using Claude 4 Sonnet and GPT-5 from the official SWE-bench Verified archive~\cite{swebenchexperiments}, with token counts estimated from public trajectories.}
\label{fig:solvecost}
\end{figure}

The pattern also changes when the harness is held fixed. Under SWE-agent, the loop/budget share is 95\% with Qwen3-Coder, 35\% with GLM-4.7-F, and 25\% with Devstral-S. AutoCodeRover shows corresponding shares of 64\%, 43\%, and 11\%. Thus, the prominence of loop/budget failures in these harnesses' aggregate profiles depends strongly on which model they are paired with.

This pairing dependence extends beyond loop and budget failures. For Devstral-S, insufficient validation accounts for 41\% of Agentless failures, 24\% under OpenHands, and 12\% under mini-SWE-agent. Yet mini-SWE-agent has a much higher loop/budget share than OpenHands, at 50\% versus 7\%, and leaves more tasks unresolved, at 78 versus 46. Its lower validation-failure share therefore accompanies a different failure profile, not better overall performance. These results highlight the need to assess the complete pairing's failure profile rather than individual categories in isolation.

Together, these comparisons show that neither a model's nor a harness's aggregate profile fully characterizes the failures of its individual pairings. The specific combination must be examined to determine which mechanisms are most prominent.

\begin{findingbox}
\textbf{Finding 4.} Harness--model pairings exhibit distinct failure profiles that are not fully characterized by either component alone. Loop or budget failures range from 3\% to 95\% across Qwen3-Coder pairings and from 25\% to 95\% across models paired with SWE-agent; insufficient validation ranges from 4\% to 41\% across Devstral-S pairings. Pairing can therefore attenuate or amplify a model-level tendency and redistribute failures among mechanisms.
\end{findingbox}

\section{RQ3: Pairing Efficiency}\label{sec:cost}

Table~\ref{tab:cost} summarizes model-call counts and charged tokens per resolved task for each harness--model pairing. Charged tokens measure token volume, not monetary cost. Pooling tokens and resolved-task counts across harnesses, Devstral-S has the lowest aggregate expenditure at 5.1 million tokens per resolved task, followed by GLM-4.7-F at 6.0 million and Qwen3-Coder at 19.8 million.

These model-level differences recur across most of the five iterative harnesses. Qwen3-Coder incurs the highest token expenditure per resolved task in four, while Devstral-S has the lowest or joint-lowest expenditure in four at the reported precision. Under SWE-agent, for example, the corresponding values are 87.8 million for Qwen3-Coder, 11.0 million for GLM-4.7-F, and 6.7 million for Devstral-S. The main exception is mini-SWE-agent, where GLM-4.7-F has the lowest expenditure at 6.2 million, followed by Qwen3-Coder at 9.5 million and Devstral-S at 9.9 million.

Model-call counts show a related pattern: Devstral-S has the lowest median under all five iterative harnesses, whereas Qwen3-Coder has the highest under four. However, fewer calls do not necessarily translate into greater token efficiency. Under mini-SWE-agent, Devstral-S has a lower median call count than GLM-4.7-F, at 62 versus 92, but consumes more tokens per resolved task, at 9.9 versus 6.2 million. Token efficiency therefore depends on the volume of tokens exchanged in those calls and the number of tasks ultimately resolved, not simply the number of calls.

Harness choice further changes expenditure within the same model. For Qwen3-Coder, moving from AutoCodeRover to SWE-agent increases median calls from 50 to 169 and tokens per resolved task from 4.8 to 87.8 million. These differences show that recurring model-level tendencies coexist with substantial harness-dependent variation: a model's typical interaction length provides useful context, but the resource requirements of a deployment must be assessed at the pairing level.

\begin{findingbox}
\textbf{Finding 5.} Among iterative harnesses, models exhibit a consistent trajectory-length tendency: Devstral-S requires the fewest calls under all five harnesses, whereas Qwen3-Coder requires the most under four. Nevertheless, harness choice changes median calls by 2.6--3.4$\times$ and charged tokens per resolved task by 9--18$\times$ within the same model. Model-level efficiency tendencies therefore coexist with substantial pairing-specific variation.
\end{findingbox}

Figure~\ref{fig:solvecost} places the evaluated pairings alongside two external OpenHands-based frontier references from the official SWE-bench Verified leaderboard archive~\cite{swebenchexperiments}, matched to the same 100 tasks. OpenHands with Claude~4~Sonnet resolves 68 tasks at an estimated 2.8 million tokens per resolved task, while OpenHands with GPT-5 resolves 70 at an estimated 1.0 million. We estimate completion tokens from assistant-message text and tool-call names and arguments, and prompt tokens by replaying preceding content at each assistant turn, using four characters per token. For each reference, tokens per resolved task are calculated by dividing the estimated total across all 100 runs by the number of tasks resolved.

Using these estimates as reference points, we find that 13 of the 18 evaluated lightweight-model pairings require more tokens per resolved task than both frontier references. Long trajectories, large per-call contexts, and expenditure on unsuccessful attempts can all increase this measure. Because frontier token counts are estimated and deployment configurations differ, the comparison is indicative and concerns token efficiency rather than monetary or computational deployment cost.

\begin{findingbox}
\textbf{Finding 6.} Among iterative harnesses, charged tokens per resolved task vary nearly 80-fold across pairings, from 1.1 to 87.8 million, and higher expenditure does not consistently yield higher resolution accuracy. Overall, 13 of the 18 evaluated pairings (72\%) resolve fewer tasks while consuming more tokens per resolved task than both estimated frontier references. Token efficiency therefore depends on how effectively the complete pairing converts inference into resolved tasks.
\end{findingbox}

\section{Discussion}\label{sec:discussion}

\subsection{Implications for Coding-Agent Deployment}\label{sec:implications}

Drawing on the observed differences across harness--model pairings, we offer empirically grounded guidance for teams selecting and operating self-hosted coding agents and for developers improving these systems.

\noindent\textbf{For application teams selecting a lightweight coding-agent stack: compare alternative harness--model pairings to guide component selection.}
A model that performs poorly under one harness may perform substantially
better under another; conversely, a leading harness does not retain its
advantage across all models. Our observed same-model resolution gap of up
to 44 percentage points shows how strongly harness choice can affect the
assessment of a model. Harnesses can attenuate or amplify model weaknesses,
while even strong pairings resolve only partially overlapping task sets.
Teams should therefore keep harness alternatives open when selecting a
model, and model alternatives open when selecting a harness. Final
selection should consider both resolution performance and the tasks
resolved on the intended workload. Neither a weak result nor a leading
score should be treated as an intrinsic property of either component: the
deployment advantage belongs to the evaluated combination of model,
harness, and tasks.

\noindent\textbf{For operations teams managing execution budgets: lightweight does not necessarily mean token-efficient.}
A smaller model's deployment footprint does not determine the inference
expenditure needed to resolve a task. Among iterative harnesses, charged
tokens per resolved task range from 1.1 to 87.8 million---nearly an 80-fold
difference. Repeated attempts, growing context, and unsuccessful runs can
consume substantial resources without producing additional resolved tasks.
Indeed, 13 of the 18 evaluated pairings consume more tokens per resolved
task than both frontier references in Figure~\ref{fig:solvecost}. Although
estimated frontier usage and differing configurations preclude conclusions
about monetary or computational cost, this comparison challenges the
assumption that smaller models necessarily require fewer tokens to
accomplish the same work. Operations teams should budget against resolved
tasks, account explicitly for expenditure on failed runs, and test whether
larger execution budgets actually yield additional resolutions. More
inference should be treated as an investment to justify, not evidence of
productive progress.

\begin{samepage}
\noindent\textbf{For coding-agent developers improving harnesses: use pairing-specific failure evidence to prioritize engineering work.} Our failure analysis shows that a mechanism prominent in a model's aggregate profile can vary substantially across harnesses. Aggregate resolution rates identify weaker configurations, but pairing-specific trajectories and failure profiles help identify which parts of the workflow warrant investigation. No-progress loops or budget exhaustion suggest examining progress-monitoring and recovery policies. Fault mislocalization points to retrieval and localization support, while insufficient validation motivates checks that exercise the required behaviour. These findings identify potential engineering priorities, not verified remedies. Developers should use the recorded evidence to investigate both model behaviour and harness--model interaction, and evaluate whether targeted changes improve the intended pairing.
\par
\end{samepage}

\subsection{Threats to Validity}\label{sec:threats}

\subsubsection{Internal and Construct Validity}

We retain a single final attempt per pairing--task combination, so run-to-run variance cannot be estimated. Although each model's sampling settings are fixed across harnesses, serving and execution variability may still affect outcomes. For each model, coverage unions combine six attempts per task, one under each harness. They therefore neither distinguish harness complementarity from run-to-run variability nor demonstrate an advantage over six attempts with a single harness. Accordingly, we emphasize broad patterns rather than small differences. Harness-specific execution limits and timeouts may also influence resolution, failure profiles, and token expenditure, limiting attribution to harness design alone.

Outcome measurement and failure labelling introduce further limitations. SWE-bench provides reproducible grading but does not assess maintainability or broader patch quality, nor establish correctness beyond the tested behaviour. Failure labelling also involves interpretation, particularly for insufficient validation and wrong or incomplete fixes. Independent model review does not eliminate this uncertainty, which may affect both individual labels and aggregate failure profiles.

\subsubsection{External Validity}

Our evaluation covers six open-source harnesses, three self-hosted coding models, and a fixed, stratified 100-task subset of SWE-bench Verified. The reported resolution rates describe this subset rather than estimate full-benchmark performance. Rankings and effect sizes may differ across systems or versions, non-Python projects, task distributions, and activities beyond repository-level issue resolution.

Generalisability also depends on serving conditions. We served all models through vLLM on a single NVIDIA H100 NVL GPU. Different sampling settings, quantization schemes, and context limits may alter resolution performance and failure profiles, while changes in caching, serving engines, or hardware may affect efficiency and outcomes for runs constrained by timeouts. We also evaluated only one completion-token threshold; different thresholds may change the observed results.

\section{Related Work}\label{sec:related}
\noindent\textbf{Coding-agent evaluation and trajectory analysis.}
Evaluations using SWE-bench~\cite{swebench} and SWE-bench
Verified~\cite{swebenchverified}, including official
submissions~\cite{swebenchexperiments}, model reports~\cite{devstral2507},
and harness studies~\cite{sweagent}, characterize the performance of selected
harness--model configurations. These results establish what the evaluated
systems achieve, but do not by themselves reveal how performance changes
when either component is replaced. Our full-factorial evaluation makes
these comparisons explicit: each model is evaluated across six harnesses,
and each harness across three models under common conditions. The resulting
differences in accuracy and resolved-task coverage show which
component-level strengths persist and which depend on the pairing.

Trajectory studies provide complementary explanations through
thought--action--result patterns~\cite{bouzeniatrajectories} and
repair-process analysis~\cite{traceability}. They identify recurring
behaviours and failure mechanisms within the studied configurations.
Building on these analytical perspectives, we compare failure profiles
across the pairing matrix rather than characterize each system in
isolation. Our results show that recognizable model-level failure
tendencies can coexist with substantial harness-dependent variation. This
extends behavioural analysis from identifying how agents fail to examining
how those failures change with the model--harness combination.

\noindent\textbf{Harness adaptation and agent evolution.}
Automated adaptation studies demonstrate that harness changes can improve
performance without replacing the underlying model. Yang
et~al.~\cite{betterharnesses} show this potential for smaller models on
business tasks, while Meta-Harness~\cite{metaharness} and Agentic Harness
Engineering~\cite{agenticharnessengineering} demonstrate improvements
through feedback-guided harness optimization. These findings align with our
observation that harness choice materially affects realized model
capability. Our contribution is complementary: the controlled pairing
matrix reveals where existing harnesses differ in task coverage, failure
profiles, and token efficiency. These comparisons provide evidence for
prioritizing model-specific harness improvements and inform model--harness
co-evolution toward more effective and efficient coding agents.

\section{Conclusion and Future Work}\label{sec:conclusion}
Our evaluation of 18 harness--model pairings shows that coding-agent performance depends on the pairing rather than either component alone. For the same model, harness choice changes resolution accuracy by up to 44 percentage points; no harness performs best across all models, and even strong pairings differ in which tasks they resolve. Failure mechanisms are similarly pairing-specific: a model's dominant failure tendency can be amplified, attenuated, or redirected by the harness. Efficiency also varies substantially: among iterative configurations using the same model, harness choice changes model calls by 2.6--3.4$\times$ and charged tokens per resolved task by 9--18$\times$. These findings caution against treating a weak or leading result as an intrinsic property of a model or harness, or assuming that lightweight models are token-efficient. Practitioners should keep both model and harness alternatives open when selecting for the intended workload, account for expenditure on unsuccessful runs when budgeting for resolved tasks, and use pairing-specific failure evidence to prioritize harness improvements.

Building on this system-level perspective, future work should develop methods for diagnosing harness--model interactions and improving their joint performance. Resolution rates and trajectory analyses show how a coding agent performs, but offer limited guidance on which interactions between model capabilities and harness policies drive success, failure, and token expenditure. A systematic framework could profile both components and use controlled interventions to identify the underlying mechanisms, guiding pairing selection and adaptation for specific tasks and resource budgets. In the longer term, these diagnoses could support a co-evolution loop: adapt the harness or train the model, evaluate the revised pairing, and use the resulting evidence to guide the next change.

\begin{samepage}
\section*{Data Availability}
We release experimental records and supporting materials for all 1{,}800 runs, including trajectories, patches, grading outcomes, failure annotations, review and configuration records, token-usage logs, and analysis scripts, through \url{https://github.com/lisazhao-gz/pairing-matters}.
\par
\end{samepage}

\IEEEtriggeratref{16}
\bibliographystyle{IEEEtran}
\bibliography{refs}

\end{document}